\documentclass[aps,pra,groupedaddress,reprint]{revtex4-2}

\usepackage[english]{babel}

\usepackage{amsmath,amssymb,amsthm,mathrsfs,bm}
\usepackage{graphicx}
\usepackage{physics}
\usepackage{soul}
\usepackage{xcolor}
\usepackage{appendix}
\usepackage{epstopdf}
\usepackage{comment}

\usepackage[colorlinks=true]{hyperref}

\begin{document}

\title{Geometry-Enabled Radiation from Structured Paraxial Electrons}

\author{M.S. Epov}%
\affiliation{School of Physics and Engineering,
ITMO University, St. Petersburg, Russia 197101}%

\author{I.E. Shenderovich}%
\affiliation{School of Physics and Engineering,
ITMO University, St. Petersburg, Russia 197101}%

\author{S.S. Baturin}%
\email{s.s.baturin@gmail.com}%
\affiliation{School of Physics and Engineering,
ITMO University, St. Petersburg, Russia 197101}%
\date{\today}

\begin{abstract}
We present a microscopic calculation of spontaneous photon emission by twisted (paraxial) electrons propagating through inhomogeneous, axisymmetric magnetic fields. We construct exact electron states that incorporate transverse mode structure and wavefront curvature by combining the Foldy-Wouthuysen transformation with a geometric framework based on Lewis-Ermakov invariants and metaplectic transformations. We show that the evolution of such structured states corresponds to an open path in the space of quadratic forms, giving rise to a geometric contribution to the emission amplitude that cannot be eliminated by gauge choice or adiabatic arguments. The inverse radius of curvature of the electron wavefront controls a finite-window boundary contribution to the emission amplitude in regions where the external magnetic field vanishes locally, this contribution vanishes when the observation window is extended to infinity. Presented mechanism generalizes Landau-level radiation to nonasymptotic, structured electron states and establishes a direct connection between noncyclic geometric evolution and photon emission.
\end{abstract}

\maketitle


 Recent theoretical and experimental studies have established that free electrons, as well as electrons propagating in static magnetic fields, may carry a well-defined projection of orbital angular momentum (OAM)~\cite{Bliokh_2017, Ivanov_2022, Alexeyev_2006, Bliokh_2006}. These so-called twisted electron states exhibit nontrivial transverse structure and phase singularities, leading to a range of novel quantum and dynamical effects. Their fundamental properties have been extensively explored~\cite{Bliokh_2007,Bliokh_2014, Schachinger_2015, Wang_2021, Cozzolino_2019, Grillo_2017, Zaytsev_2017, Silenko_2025,Ivanov1,Ivanov2}, yet several open questions remain, including the stability of vortex states and their behavior under realistic external-field configurations~\cite{Ivanov_2022}.

Quantum electrodynamical processes involving structured particles - most notably electrons prepared in Laguerre-Gaussian-type modes~\cite{Allen_1992} - have primarily been studied for stationary states and uniform external fields \cite{Karlovets_2023}, often within the dipole approximation. In such settings, photon emission is conventionally attributed to local acceleration induced by external forces. However, this picture becomes incomplete for nonasymptotic, structured wave packets whose evolution is governed by geometric phases and intrinsic wavefront curvature.

In this work, we develop a theoretical framework for spontaneous photon emission by twisted electrons propagating in a general inhomogeneous axisymmetric magnetic field. Starting from a relativistic description, we apply the Foldy-Wouthuysen transformation~\cite{Foldy_1950} followed by a controlled paraxial reduction, which allows the longitudinal coordinate to play the role of an effective evolution parameter. The resulting transverse dynamics is governed by a Lewis-type system with a spatially dependent frequency. By exploiting the Lewis-Riesenfeld invariant theory~\cite{Lewis1,Lewis2,Lewis} and the Ermakov transformation~\cite{Aldaya_2011,Guerrero_2014,Enk_2021,Filina_2023,Filina_2024,Filina_2025}, we obtain an exact representation of the electron states in terms of metaplectic transformations acting on harmonic-oscillator eigenstates. This construction provides a nonperturbative and fully geometric description of twisted electron beams in inhomogeneous magnetic fields.

\begin{figure}[b]
    \centering
    \includegraphics[width=1\linewidth]{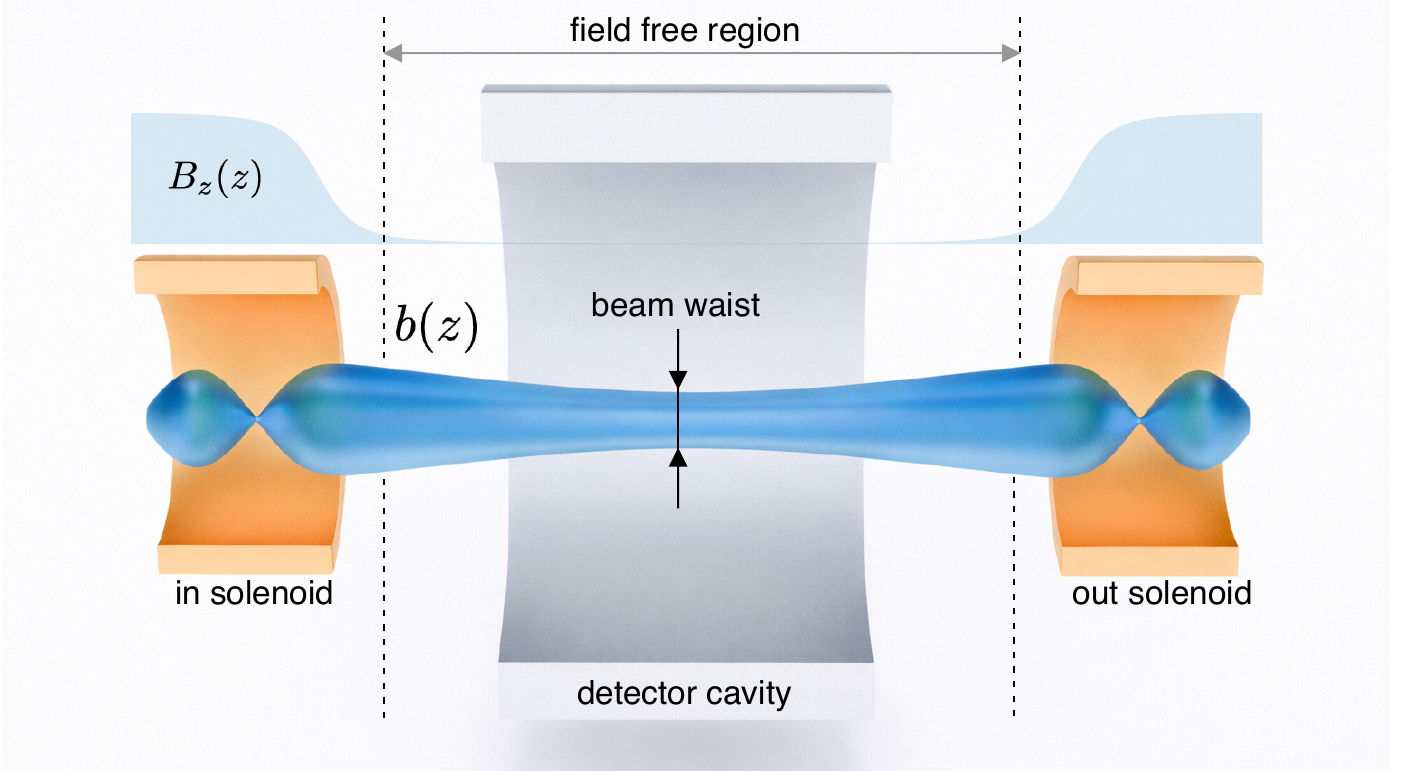}
    \caption{Schematic setup: two solenoids shape an axisymmetric guiding field $B_z(z)$ and imprint a nontrivial Ermakov scaling $b(z)$ (state transverse size and wavefront curvature) on the paraxial twisted-electron mode. Photon detection occurs in the central cavity located in a locally field-free region, where radiation persists due to the inherited curvature.}
    \label{fig:emission}
\end{figure}

Our central result is the identification of an irremovable geometric contribution to the photon emission amplitude associated with the inverse radius of curvature of the electron wavefront. This contribution originates from a quadratic phase generated by the metaplectic shear inherent to paraxial evolution and is uniquely fixed by the Ermakov equation. The quadratic wavefront phase is not an electromagnetic gauge artifact. It is a state-dependent phase-space shear fixed by the Ermakov transport, and it enters the photon-emission matrix element through the prepared transverse state. 
As a consequence, photon emission can persist in spatial regions where the external magnetic field locally vanishes, provided the electron wave packet retains finite curvature inherited from its prior evolution in the field. The resulting emission process probes a reduced geometric memory of the focusing transport. The field-free amplitude is governed by the prepared transverse envelope and phase structure, together with the finite observation window, rather than by the instantaneous local field strength alone (see Fig.~\ref{fig:emission}). Although the radiation in the field-free region is the radiation of a freely propagating structured beam, in the proposed magnetic-transport setup the relevant structural parameters are fixed by the upstream solenoidal evolution.
We emphasize that the field-free contribution is a finite-window boundary effect. The finite observation boundaries prevent complete phase cancellation, while the inherited state curvature controls the surviving amplitude. The contribution vanishes when the observation window is extended to infinity.

More broadly, our results generalize radiation mechanisms associated with Landau-level transitions to structured, nonasymptotic electron states and establish a direct connection between geometric dressing of matter waves, photon emission, and concepts familiar from cavity QED \cite{Kimble1998}, where squeezing-induced geometric phases are known to modify light-matter coupling strengths.

Throughout this paper, we adopt natural units with $\hbar=c=1$ and assume $e<0$ for the electron charge.


The quantum-mechanical description of relativistic electrons in external electromagnetic fields exhibits a structure closely related to its nonrelativistic counterpart. The dynamics of a relativistic electron in a stationary magnetic field are governed by the Dirac equation \cite{Dirac_1928}
\begin{align}
\left(\boldsymbol{\alpha \hat{\pi}}+\beta m\right)\Psi=i\partial_t \Psi,
\end{align}
where $\hat{\boldsymbol{\pi}}=\hat{\mathbf{p}}+|e|\mathbf{A}_{cl}$ is the kinetic momentum operator and $\Psi(\vec{r},t)=\Psi_{st}(\vec{r})e^{-iEt}$ is a bispinor with fixed energy $E$.

We consider an axisymmetric magnetic field predominantly aligned along the $z$ direction,
\begin{align}
\mathbf{B}(z)^T=[-x B_z'(z)/2,-y B_z'(z)/2,B_z(z)].
\end{align}
The transverse magnetic components are uniquely fixed by the solenoidal condition $\nabla\cdot\mathbf{B}=0$ and therefore cannot be neglected. In the Coulomb gauge, the corresponding vector potential reads \cite{Filina_2024}
\begin{align}
\mathbf{A}^T_{cl} = [-y B_z(z)/2,x B_z(z)/2,0],
\end{align}
which automatically incorporates the transverse magnetic field. Alternative gauge choices redistribute longitudinal and transverse couplings but do not modify the underlying paraxial dynamics (see Ref.~\cite{Filina_2025}).

Squaring the stationary Dirac equation for a purely magnetic field (or, equivalently, squaring the exact Foldy-Wouthuysen Hamiltonian~\cite{Silenko_2025}) yields a Pauli-type equation for the upper spinor $\Phi_{FW}$,
\begin{align}
\left(\hat{\pi}^2 - e \mathbf{B}\cdot\boldsymbol{\sigma}\right)\Phi_{FW}=k^2 \Phi_{FW}, \qquad k^2=E^2-m^2.
\end{align}
Note that in the absence of an electric field, the standard Foldy-Wouthuysen reduction to this order does not produce spin-orbit or Darwin terms.

The second term represents the Pauli spin-field interaction. In the regime considered here, the transverse spin-dependent couplings are parametrically small ($\propto 1/(k \rho_H^2)$), while the longitudinal term is diagonal in $s_z$ and produces only a small spin-dependent shift of the effective level structure. We therefore work in the spin-conserving sector, retain only transitions with $\Delta s_z=0$, and neglect spin-flip channels, which lie beyond the accuracy of the present scalar/paraxial approximation.

In the paraxial regime $p_z\gg p_\perp$, we write $\Phi_{FW}(\vec{r},t)=(\psi(\vec{r}),0)^T\exp(ikz-iEt)$, where the envelope $\psi$ varies slowly along $z$. This leads to the scalar paraxial equation
\begin{align}
\hat{\pi}_{\perp}^2\psi=2ik\partial_z\psi,
\label{eq:Fin_Ham}
\end{align}
which has the form of a Schr\"odinger equation with the propagation coordinate $z$ playing the role of time.

Expanding the kinetic momentum operator, Eq.~\eqref{eq:Fin_Ham} can be rewritten as
\begin{align}
\label{eq:oscHfull}
&i\frac{\partial \psi}{\partial z}
=\left[\hat{\mathcal{H}}_\perp(z)+\Omega(z)\hat{L}_z\right]\psi, \\
\label{eq:oscH}
&\hat{\mathcal{H}}_\perp(z)=\frac{\hat{p}_\perp^2}{2} + \frac{\Omega^2(z)\hat{\rho}^2}{2},
\end{align}
where $\hat{L}_z$ is the orbital angular momentum operator and $\hat{\rho}^2=\hat{x}^2+\hat{y}^2$. We introduce normalized coordinates $\tilde{z}=z/(k\rho_H^2)$, $\tilde{x}=x/\rho_H$, and $\tilde{y}=y/\rho_H$, where the magnetic length and normalized frequency are defined as
\begin{align}
\rho_H=\sqrt{\frac{2}{|e|\max |B_z(z)|}},\qquad \Omega(z)=\frac{B_z(z)}{\max |B_z(z)|}.
\end{align}
For clarity, we drop tildes in the following and work exclusively with normalized variables.

The Larmor rotation operator
\begin{align}
\hat{R}=\exp \left[-i\int_0^z d\bar z\,\Omega(\bar z)\hat{L}_z\right]
\end{align}
can be factored out exactly, since $\hat{\mathcal{H}}_\perp(z)$ commutes with $\Omega(z)\hat{L}_z$ for all $z$. This commutation follows directly from the rotational symmetry of the transverse harmonic confinement. The remaining dynamics generated by $\hat{\mathcal{H}}_\perp(z)$ together with $i\partial_z$ constitutes a Lewis-type system: a two-dimensional harmonic oscillator with $z$-dependent frequency, where the propagation coordinate $z$ replaces time.

This two-dimensional system decomposes into two equivalent one-dimensional subsystems, each admitting an exact invariant of motion,
\begin{align}
\label{eq:ints}
\hat{I}_x=\frac{(b\hat{p}_x-b'x)^2}{2}+\frac{\hat{x}^2}{2b^2}, \\
\hat{I}_y=\frac{(b\hat{p}_y-b'y)^2}{2}+\frac{\hat{y}^2}{2b^2}.
\end{align}
Here $b(z)$ satisfies the Ermakov equation
\begin{align}
b''+\Omega^2(z)b=\frac{1}{b^3}, \qquad
b(0)=b_0,\quad b'(0)=b'_0.
\end{align}
The Ermakov function $b(z)$ is uniquely fixed by the field profile $\Omega(z)$ and the state parameters at the reference plane $z=0$.

The invariants \eqref{eq:ints} are unitarily equivalent to the Hamiltonian of a unit-frequency harmonic oscillator via a composition of metaplectic transformations,
\begin{align}
\hat{H}_0=\hat{\mathrm{S}}^{\dagger}\hat{\mathrm{M}}^{\dagger}(\hat{I}_x+\hat{I}_y)\hat{\mathrm{M}}\hat{\mathrm{S}},
\end{align}
where
\begin{align}
\label{eq:HO}
&\hat{H}_0=\frac{\hat{p}_\perp^2}{2}+\frac{\hat{\rho}^2}{2}, \\
&\hat{\mathrm{S}}=\exp\!\left[-i \frac{\ln b}{2}\left(\hat{\bm{\rho}}\cdot\hat{\mathbf{p}}_\perp+\hat{\mathbf{p}}_\perp\cdot\hat{\bm{\rho}}\right)\right], \\
&\hat{\mathrm{M}}=\exp\!\left[i \frac{b'}{b} \frac{\hat\rho^2}{2}\right].
\end{align}
The Lewis-Ermakov invariants thus serve a dual role: they label the transverse modes and simultaneously encode the canonical transformation that renders the $z$-dependent dynamics stationary. Indeed, the structure of $\hat{I}_{x,y}$ suggests introducing new phase-space variables $(\hat{Q}_{x,y},\hat{P}_{x,y})$ defined by $Q_j=x_j/b$ and $P_j=b\,\hat{p}_j-b'x_j$, in terms of which each invariant takes the standard harmonic-oscillator form $\hat{I}_j=\tfrac{1}{2}(\hat{P}_j^2+\hat{Q}_j^2)$. The operators $\hat{\mathrm{S}}$ and $\hat{\mathrm{M}}$ implement the corresponding symplectic transformation at the quantum level, providing its metaplectic lift \cite{Folland1989}.

The full solution of Eq.~\eqref{eq:oscHfull} can therefore be written as a squeezed number state (for the relevant derivations see Ref.\cite{Fernandez2003})
\begin{align}
\label{eq:psi}
\psi=\hat{\mathrm{R}}\hat{\mathrm{U}}\hat{\mathrm{M}}\hat{\mathrm{S}}|n_+,n_-\rangle,
\end{align}
where $|n_+,n_-\rangle$ is a fixed Fock state of the two-dimensional harmonic oscillator and $n_+$ and $n_-$ are the left and right circular numbers. In the polar configuration space the state has the form of a Laguerre-Gaussian (twisted) mode \cite{SpatMode}
\begin{align}
    &\langle \rho, \varphi| n_+,n_- \rangle= \\ \nonumber
    &\sqrt{\frac{n_r!}{\pi(n_r+|l|)!}} \rho^{|l|}\exp\left(-\frac{\rho^2}{2}\right)
    \mathcal{L}_{n_r}^{|l|}(\rho^2)\exp\left(il\varphi\right), \\
    \nonumber ~~~~~~~~~&n_r=\mathrm{min}(n_+,n_-), ~~\ell=n_+-n_- \nonumber
\end{align}
Above $\mathcal{L}$ is the generalized Laguerre polynomial and we imply standard normalization for the generalized ladder operators 
\begin{align}
    \hat{a}_{\pm}=\frac{\hat{a}_x\mp i\hat{a}_y}{\sqrt{2}},~~[\hat{a}_{\pm},\hat{a}_{\pm}^\dagger]=1.
\end{align}

The operator
\begin{align}
\hat{\mathrm{U}}=\exp\!\left[-i{(n_++n_-+1)}\int_0^z\frac{d\bar z}{b^2(\bar z)}\right]
\end{align}
contains the Lewis phase accumulated along the propagation.

Geometrically, the Lewis phase represents the holonomy of the connection induced by the metaplectic representation on the bundle of invariant eigenstates. For cyclic evolution it reduces to a purely geometric phase determined by the enclosed curvature, while for open trajectories relevant to realistic structured electron beams it acquires boundary contributions that cannot be removed by gauge choice. Unlike a conventional gauge phase, this quadratic (metaplectic) phase is fixed once the Lewis-Ermakov invariant structure and boundary conditions are specified; it cannot be removed by a rephasing that preserves the invariant eigenbasis and the associated symplectic (metaplectic) map. It  represents the geometric dressing of the electron wavefunction by the external magnetic field, encoding the cumulative effect of transverse focusing into a history-dependent phase. Physically, this implies that the electron propagates not as a free particle, but as a ``dressed'' mode that dynamically adapts its wavefront curvature to the profile of the guiding magnetic potential.

Finally, we emphasize that the ordering of the metaplectic operators in Eq.~\eqref{eq:psi} is fixed by the structure of the invariants \eqref{eq:ints}. 

Now we focus on the spontaneous emission process $e_i \rightarrow e_f + \gamma$ for a relativistic twisted electron propagating along the $z$-axis in an inhomogeneous magnetic field. The theoretical framework is based on scalar QED, where the photon is a plane-wave mode and the electron wavefunction is given by a paraxial state in a non-uniform magnetic 
\begin{align}
    \Psi=\frac{1}{\sqrt{L}}\psi\exp\left[i\chi^2 z-iEt \right].
\end{align}
Here the transverse part $\psi$ is defined in Eq.\eqref{eq:psi}, $k$ as before the total wave vector of the electron, $E$ is the total energy and $L$ is the finite $z$-length of the considered volume measured in the units of $k\rho_H^2$ and 
\begin{align}
\chi=k\rho_H.    
\end{align}
The first-order transition amplitude is ~\cite{SokolovTernov1974}
\begin{align}
&S_{fi} = i  \int\limits_{-\infty}^{\infty} dt \int\limits_{-\infty}^{\infty} dx\int\limits_{-\infty}^{\infty} dy \int\limits_{-L/2}^{L/2}dz\, \boldsymbol{j}_{fi}\cdot \mathbf{A}^*_{ph}, \\
&\boldsymbol{j}_{fi}=\frac{e\beta}{2\chi}\left[\Psi_f^*(\hat{\boldsymbol{\pi}} \Psi_i)+\Psi_i (\hat{\boldsymbol{\pi}}\Psi_f)^* \right]
\label{eq:Sfi}
\end{align}
where $\mathbf{A}_{ph} = \frac{\boldsymbol{\epsilon}}{\sqrt{2\omega V}}\exp(-i\omega t) \exp(i\rho_H\mathbf{k}^{\omega}_\perp\cdot \boldsymbol{\rho}+i k\rho_H^2 k_z^{\omega} z)$ is the photon wave function with the photon wave vector $\mathbf{k}^\omega=\omega (\sin\theta \cos\varphi, \sin\theta \sin\varphi, \cos\theta)^T$ and polarization vector $\boldsymbol{\epsilon}$ satisfying the transversality condition $\boldsymbol{\epsilon}^* \cdot \mathbf{k}^{\omega} = 0$ and $\beta=k/E$ is the electron relativistic beta-factor. 

We note that as before all spatial coordinates are dimensionless, the kinetic momentum operator is dimensionless as well and normalized to $1/\rho_H$.
We integrate over the time and get
\begin{align}
\label{eq:Amp02}
   & S_{fi}=\int\limits_{-L/2}^{L/2} dz\frac{\pi ie\beta}{\chi L \sqrt{2\omega V}}\delta(E_f+\omega-E_i) 
   \times \\  & \boldsymbol{\epsilon}^*\cdot\left[\int \Psi_f^* \left\{\exp(-i \boldsymbol{\varkappa\cdot\hat{\mathbf{r}}}) \hat{\boldsymbol{\pi}}\right\}_+ \Psi_i  dxdy\right].\nonumber
\end{align}
Above 
$\boldsymbol{\varkappa}=\rho_H( k_x^\omega, k_y^\omega,\chi k_z^\omega)^T$, $\{\cdot ,\cdot\}_+$ is the anticommutator and $\hat{\boldsymbol{\pi}}=(\hat{p}_x-\Omega(z)\hat{y},\hat{p}_y+\Omega(z)x, \hat{p}_z/\chi)^T$
We note that
\begin{align}
    \left\{\exp(-i \boldsymbol{\varkappa\cdot\hat{\mathbf{r}}}) ,\hat{\boldsymbol{\pi}}\right\}_+=\exp(-i \boldsymbol{\varkappa\cdot\hat{\mathbf{r}}}) [ 2\hat{\boldsymbol{\pi}}-\rho_H\mathbf{k}^\omega]
\end{align}
switch to the Fock representation, and rewrite Eq.\eqref{eq:Amp02} in a slightly modified form
\begin{align}
\label{eq:Amp2}
    &S_{fi}= \\ \nonumber
    &\int\limits_{-L/2}^{L/2}
    \frac{2 \pi i e \beta}{\chi L \sqrt{2\omega V}} 
\left[s_\perp(z)+s_\parallel(z)\right]\delta(E_f+\omega-E_i) dz.
\end{align}
with 
\begin{align}
\label{eq:sperp_raw}
    s_\perp(z)=&\exp[i\delta \phi(z)] \times \\ 
    &\langle n^f_+,n^f_-| \hat{\mathrm{S}}^{\dagger}\hat{\mathrm{M}}^{\dagger} \boldsymbol{\epsilon}^*_\perp\cdot\hat{\mathbf{J}}_\perp\hat{\mathrm{M}}\hat{\mathrm{S}}| n^i_+,n^i_- \rangle.\nonumber
\end{align}
Above the operator $\hat{\mathbf{J}}_\perp$ is given by
\begin{align}
\hat{\mathbf{J}}_\perp=\exp(-i \boldsymbol{\varkappa}_\perp \cdot \hat{\boldsymbol{\rho}})\hat{\boldsymbol{\pi}}_\perp.  
\end{align}
The function $s_\parallel(z)$ in the second term of the Eq.\eqref{eq:Amp2} is given by
\begin{align}
    s_\parallel(z)=&\frac{\exp[i\delta \phi(z)]}{\chi} 
    \langle n^f_+,n^f_-| \hat{\mathrm{S}}^{\dagger}\hat{\mathrm{M}}^{\dagger} \epsilon_z^*\hat{\mathrm{J}}_z\hat{\mathrm{M}}\hat{\mathrm{S}}| n^i_+,n^i_- \rangle,
\end{align}
with 
\begin{align}
\hat{\mathrm{J}}_z=\exp(-i \boldsymbol{\varkappa}_\perp \cdot \hat{\boldsymbol{\rho}})[\hat{\mathcal{H}}_\perp(z)+\Omega(z)\hat{L}_z]. 
\end{align}
Above $\hat{\mathcal{H}}_\perp(z)$ is given by Eq.\eqref{eq:oscH} and we have used the identity between the right and the left hand side of Eq.\eqref{eq:oscHfull}.
The common phase factor is given by
\begin{align}
\label{eq:LLphase}
    \delta \phi(z)&=\frac{\Delta k-k_z^{\omega}}{k^i} \chi^2 z-\Delta N\int\limits^z_{-L/2}\frac{d\bar z}{b^2(\bar z)}-\Delta l \int\limits^z_{-L/2} \Omega(\bar z) d\bar z. \nonumber\ \\
    &\Delta N= \Delta n_{+}+\Delta n_{-}, ~~\Delta l= \Delta n_{+}-\Delta n_{-}, \nonumber\\
    &\Delta k= k^i-k^f,~~ \chi=k^i \rho_H.
\end{align}
Here $\Delta n_\pm=n_\pm^i-n_\pm^f$.

For the common laboratory (electron microscope) setup parameter $\chi$ is significantly larger than unity. For instance for the peak magnetic field $\max |B(z)|\sim 1$ T  of the magnetic guiding system and the electron kinetic energy $W_e \sim 100$ keV we have for the characteristic magnetic length $\rho_H=36.3$ nm and for the modulus of the electron wave vector $k=1.7$ pm$^{-1}$. This results in $\chi\approx 6.2 \times 10^4$. To proceed further we assume $\chi$ to be large and neglect the $s_\parallel(z)$ contribution to the transition amplitude as it is suppressed by $1/\chi$ factor in comparison to the $s_\perp(z)$.   

To proceed we first calculate the transformation under the brackets in Eq.\eqref{eq:sperp_raw} 
\begin{align}  
\label{eq:perpcurtr}
\hat{\mathrm{S}}^{\dagger}\hat{\mathrm{M}}^{\dagger}\hat{\mathbf{J}}_\perp\hat{\mathrm{M}}\hat{\mathrm{S}}= \exp\left[-i b(z) \boldsymbol{\varkappa}_\perp \cdot \hat{\boldsymbol{\rho}}\right] \hat{\bar{\boldsymbol{\pi}}}_\perp 
\end{align}
with 
\begin{align}
\label{eq:finkinm}
    \hat{\bar{\boldsymbol{\pi}}}_\perp =\frac{\hat{\mathbf{p}}_\perp}{b(z)}+b'(z) \hat{\boldsymbol{\rho}}+b(z)\Omega(z)\mathbf{u}_z\times\hat{\boldsymbol{\rho}}
\end{align}
above $\mathbf{u}_z$ is the unit vector along the $z$-axis. In a locally field-free region, $\Omega(z)=0$, while the curvature term $b'(z)\hat{\boldsymbol{\rho}}$ remains. This term is fixed by the metaplectic shear (and hence by the Ermakov solution) and therefore acts as an effective geometric driving that enables radiation even when $B_z(z)=0$ locally.

It is convenient to rewrite operator Eq.\eqref{eq:finkinm} in a circular form 
\begin{align}    
\label{eq:circmom}
\hat{\bar{\pi}}_\sigma=\hat{\bar{\pi}}_x+\sigma i \hat{\bar{\pi}}_y
\end{align}
where $\sigma=\pm$ and denote right and left circular component. Switching to generalized ladder operators we have
\begin{align}
    \hat{\bar{\pi}}_{-\sigma}=C_{a,\sigma}\hat{a}_\sigma+C_{a^\dagger,\sigma}\hat{a}_\sigma^{\dagger},
\end{align}
The coefficients are given by.
\begin{align}
\label{eq:cfunc}
              &C_{a,\sigma}=b'(z)-\sigma i b(z)\Omega(z)-\frac{i}{b(z)}, \nonumber\\
    &C_{a^\dagger,\sigma}=b'(z)-\sigma i b(z)\Omega(z)+\frac{i}{b(z)}.
\end{align}
The exponent in Eq.\eqref{eq:perpcurtr} factorizes as 
\begin{align}
\label{eq:eqxpfact}
  &\exp\left[-i b(z) \boldsymbol{\varkappa}_\perp \cdot \hat{\boldsymbol{\rho}}\right]=\prod\limits_{\sigma=\pm}\exp\left[-i(\kappa_\sigma\hat{a}_\sigma+\kappa_\sigma^*\hat{a}_\sigma^\dagger) \right],
\end{align}
with 
\begin{align}
    \kappa_\sigma=\frac{b(z)\rho_H k_\perp^{\omega}}{2} \exp(\sigma i \varphi).
\end{align}

Because the transverse Hamiltonian Eq.\eqref{eq:HO} is a sum of two commuting 1D oscillators, the Fock space factorizes as
$|n_+,n_-\rangle=|n_+\rangle\otimes|n_-\rangle$ with
$[\hat{a}_+,\hat{a}_-]=[\hat{a}_+,\hat{a}_-^\dagger]=0$. Given the factorization Eq.\eqref{eq:eqxpfact} we note that matrix element $s_\perp(z)$ can be expressed as a product of one dimensional form factors that we define as  
\begin{align}
\label{eq:formf}
   &\mathcal{F}_{n^f_\sigma,n^i_\sigma}(\kappa_\sigma)= \langle n^f_\sigma| \hat{D}_\sigma(-i\kappa_\sigma^*)| n^i_\sigma \rangle, \\
   &\hat{D}_\sigma(-i\kappa_\sigma^*)=\exp \left[-i(\kappa_\sigma\hat{a}_\sigma+\kappa_\sigma^*\hat{a}_\sigma^\dagger) \right]. \nonumber
\end{align}
This is a common matrix element of a displacement operator $\hat{D}_\sigma$ that can be evaluated analytically from the action of the ladder operators on the number states. We sketch derivation in the 
Supplemental Material \cite{supp}
and provide the exact expression for $\mathcal{F}_{n^f_\sigma,n^i_\sigma}(\kappa_\sigma)$.

With the help of Eq.\eqref{eq:circmom} one may immediately evaluate the weighting coefficient and express it in terms of $C$ functions of Eq.\eqref{eq:cfunc} and form factors Eq.\eqref{eq:formf}  
\begin{align}
\label{eq:P_def}
&\mathcal{P}_{n^f_\sigma n^i_\sigma}(z)
\equiv
\left\langle n^f_\sigma \right|
\hat{D}_\sigma(-i\kappa_\sigma^*)\,\hat{\bar{\pi}}_{-\sigma}(z)
\left| n^i_\sigma \right\rangle
= \\
&C_{a,\sigma}\sqrt{n^i_\sigma} \mathcal{F}_{n^f_\sigma,n^i_\sigma-1}(\kappa_\sigma)
+C_{a^\dagger,\sigma}\sqrt{n^i_\sigma+1} \mathcal{F}_{n^f_\sigma,n^i_\sigma+1}(\kappa_\sigma). \nonumber
\end{align}
We note that above $\mathcal{F}$ and the coefficients $C_{a,\sigma},C_{a^\dagger,\sigma}$ depend on $b(z)$, $b'(z)$ and $\Omega(z)$, and thus inherit a parametric $z$-dependence. 
Next we use the identity
\begin{align}
\boldsymbol\epsilon_{\lambda,\perp}^*\cdot\hat{\bar{\boldsymbol\pi}}_\perp
=
\frac12\sum_{\sigma=\pm} \epsilon_{\lambda\sigma}^* \hat{\bar{\pi}}_{-\sigma}.
\end{align}
with $\lambda=\pm1$ corresponding to the photon polarization and proceed to the final expression for the $s_\perp(z)$ in a compact form
\begin{align}
\label{eq:sperp}
&s_\perp(z)=\frac{e^{i\delta\phi(z)}}{2}\sum_{\sigma=\pm}
\epsilon_{\lambda\sigma}^*
\mathcal{P}_{n^f_\sigma n^i_\sigma}
\prod_{\sigma'\neq\sigma} \mathcal{F}_{n^f_{\sigma'} n^i_{\sigma'}}(\kappa_{\sigma'}).
\end{align}
The polarization vector component is given by
\begin{align}
    \epsilon_{\lambda\sigma}=\frac{\cos \theta -\lambda \sigma}{\sqrt{2}}\exp(\sigma i \varphi).
\end{align}

Equations~\eqref{eq:sperp}, \eqref{eq:LLphase}, and \eqref{eq:cfunc} provide the general emission amplitude in an arbitrary axisymmetric inhomogeneous field. The geometry enters through the accumulated Lewis/Larmor phase given by Eq.\eqref{eq:LLphase}, through the phase of the coefficients $C_{a,\sigma}$, $C_{a^\dagger,\sigma}$ (Eq.\eqref{eq:cfunc} and the displacement parameters $\kappa_\sigma$, all determined by $b(z)$ and $b'(z)$. This dependence in a general case of inhomogeneous magnetic field enables engineered enhancement or suppression of the photon emission.

\begin{figure}[t]
    \centering
    \includegraphics[width=1\linewidth]{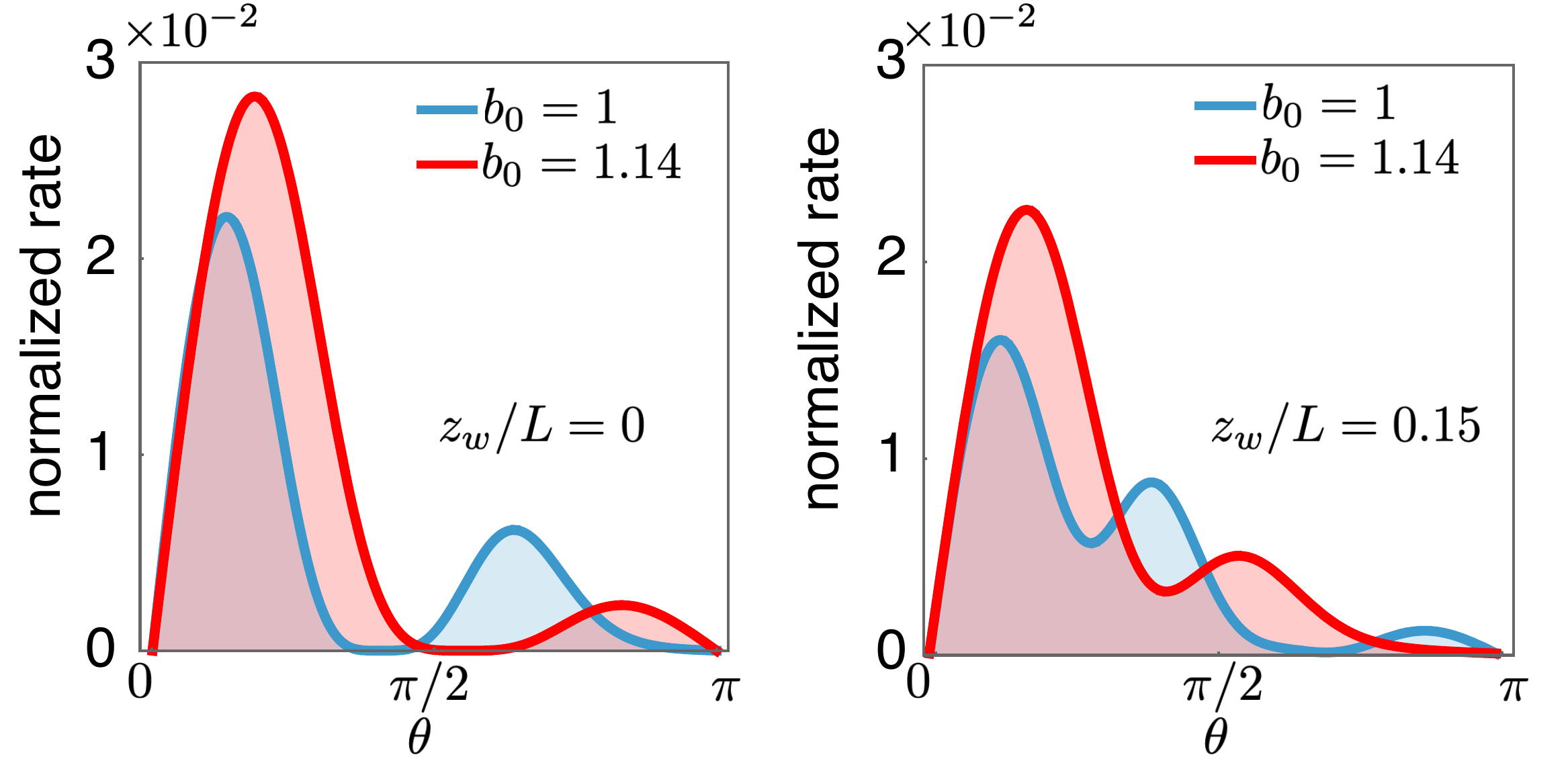}
    \caption{Normalized differential rates for the field-free region calculated for $\beta=0.548$ (electron kinetic energy $100$~keV) using Eq.~\eqref{eq:Rrate} as a function of the angle $\theta$. The left panel uses interaction length $L=30$ and $z_w/L=0$, corresponding to a waist at the center of the detector region. The right panel uses the same $L=30$ but $z_w/L=0.15$. The change in angular dependence and the slight rate drop show that the angular distribution depends on the upstream state preparation.
    } 
    \label{fig:Nrate}
\end{figure}

Next we evaluate the differential emission probability ${d\dot{w}}=dw/T=|S_{fi}|^2/T d n_{\omega}$ 
in a field-free region $\Omega=0$. 
Here $d n_{\omega}=V\omega^2 \sin\theta d\theta d\varphi /(2\pi)^3$ is the photon density of states in vacuum. With the help of Eq.\eqref{eq:Amp2} we get
\begin{align}
    d\dot{{w}}=\alpha\frac{4\pi \beta^2}{\chi^2 L^2} \frac{2\pi\delta(E_f+\omega-E_i)}{2\omega V}\left|\int\limits_{-L/2}^{L/2} s_\perp(z) dz\right|^2 dn_{\omega}
\end{align}
Above we used the identity $e^2=4\pi \alpha$ with $\alpha$ being the fine structure constant. We calculate the integral under the modulus in the dipole approximation (that is well justified by the large $\chi\gg 1$), integrate over $\omega$ and $\varphi$ angle of the photon (see 
Supplemental Material \cite{supp}
for details), sum over the photon polarizations and arrive at the expression for the differential rate for the twisted states with the initial orbital angular momentum $\ell^i$
\begin{align}
\label{eq:rate}
    \frac{d\dot{{w}}}{d\theta}=\ell^i\frac{\alpha\beta^4\gamma m_e}{\chi^4} R(\theta,L,b_0,z_w).
\end{align}
Here $b_0$ is the state size at the waist (focal point) and $z_w$ is the relative position of the waist to the center of the detector window. 
The normalized rate that we plot in Fig.\ref{fig:Nrate} for relevant parameters is given by
\begin{align}
\label{eq:Rrate}
&R(\theta,L,b_0,z_w)=
Q(X,z_w)
\frac{\left[1+(\cos\theta)^2 \right]\sin\theta}{8b_0^4}, 
\\
&Q(X,z_w)=\frac{1-2\cos\left(2X\right)\cos\left(4X \frac{z_w}{L}\right)+\cos^2\left(2X\right)}{4X^2}, \nonumber
\\
 &X=\frac{1-\beta \cos\theta}{8b_0^2}L.\nonumber
\end{align}
The structure of $R(\theta,L,b_0,z_w)$ reveals a characteristic diffractive pattern, where the interaction length $L$ acts as an effective aperture and $z_w$ parametrically controls the interference pattern. This structure is analogous to Fraunhofer diffraction from a finite slit.
We note the boundary character  in Eq.~\eqref{eq:Rrate}. Since $X\propto L$ while the numerator of $Q(X,z_w)$ remains bounded, the rate vanishes as $L\to\infty$. The field-free signal is therefore a finite-window boundary contribution.

Apart from the envelope $Q(X,z_w)$, the rate given by Eq.~\eqref{eq:rate} coincides with the standard electric-dipole (E1) Landau-transition scaling for the channel $n_r=0$ and $\Delta \ell=+1$. It is proportional to the initial Landau quantum number (i.e., $\propto \ell^{i}$), carries the usual fine-structure constant factor $\alpha$, and reproduces the expected paraxial/relativistic dependence through $\beta$, $\gamma$, and the magnetic-length scale via $\chi=k\rho_H$ (with $\rho_H^2=2/(|e|B_{\max})$ and $B_{\max}=\max|B_z(z)|$).

Experimental setup shown in Fig.~\ref{fig:emission} can be implemented in a transmission electron microscope, where vortex beams are routinely produced~\cite{Saitoh_2012,Verbeeck_2010,doi:10.1126/science.1198804,Schachinger_2015,Grillo_2017,Schattschneider_2014}. We emphasize that all rate estimates quoted below refer to a low-current, highly coherent $100\,\mathrm{nA}$ dc beam and already include the total number of electrons simultaneously present in the detector volume, assuming incoherent summation over independently radiating electrons. For $B_{\max}=1\text{--}10\,\mathrm{T}$, $W_e\simeq100\,\mathrm{keV}$, $b_0=1.14$, and $L=30$, Eq.~\eqref{eq:rate} corresponds to microwave emission in the $\nu\simeq4.5\text{--}45\,\mathrm{GHz}$ range and a field-free interaction length of $6.7\text{--}0.67\,\mathrm{cm}$. In this $100\,\mathrm{nA}$ dc-beam regime, the free-space rate integrated over $\theta$ scales as $\ell^i\times(0.25\text{--}2.5)\,\mathrm{s^{-1}}$. For the initial orbital angular momentum projection $\ell^i\simeq10^3$ (as reported in Ref.~\cite{huge_L}) and $B_{\max}\simeq10\,\mathrm{T}$, this yields a total photon count of order $2.5\times10^3\,\mathrm{s^{-1}}$, compatible with superconducting microwave readout. We note that in the low-current dc TEM regime backgrounds such as wakefield and beam-driven cavity effects are absent to leading order for a dc beam, transition radiation is absent in the idealized geometry since the electron does not cross material interfaces in the field-free region, diffraction and aperture radiation can be made parametrically small by choosing the beam radius much smaller than the aperture radius, and residual fringe radiation can be reduced by smooth magnetic matching and by placing the detector sufficiently far from the magnetic edge. Thermal microwave backgrounds can be further suppressed by cryogenic operation. A resonant cavity can further enhance the detectable count via the Purcell factor~\cite{Purcell1946}.


In conclusion, we developed a general QED framework for paraxial structured charged particles in axisymmetric, $z$-dependent external fields. A key consequence is that these states can radiate in a locally field-free region within a finite observation window. The emission amplitude retains a geometric ``memory'' of the prior evolution through the Ermakov scaling $b(z)$ and $b'(z)$, and in the appropriate limit the rate recovers Landau-like transition scaling, while the field-free contribution remains a boundary effect induced by the inherited state curvature rather than a bulk process. The phase cancellation that would suppress emission in an exactly field-free region is fragile, so weak background fields can modify the accumulated phase and may \emph{increase} the net rate. Finally, breathing (nonstationary Landau) structured states \cite{Silenko_2021,KarlovetsNJP2021} in inhomogeneous  \cite{Filina_2023,Filina_2025} can be manipulated to enhance or suppress emission by tailoring $b(z)$, providing a geometric control knob for tunable radiation from structured electron beams. Beyond the dipole channel, transitions with $\Delta \ell>1$ exhibit a saddle-point structure in the emission integral, leading to a preference for higher-order OAM-changing processes. In the paraxial large-$\chi$ regime these channels are parametrically suppressed as $\sim 1/\chi$ relative to the dipole contribution; however, when $\chi\sim\mathcal{O}(1)$ they are no longer suppressed and can dominate the dipole rate.

\begin{acknowledgments}

The authors thank Ivan Terekhov for important comments. 

\end{acknowledgments}

\bibliographystyle{apsrev4-2}
\bibliography{refs}

@misc{supp,
	note = {See Supplemental Material at [URL] for details on the (i) explicit evaluation of the matrix elements, (ii) discussion on the photon wave vector and (iii) evaluation of the emission amplitude and rate for the case of the locally field free region.}}

@article{Fernandez2003,
	author = {Fern{\'a}ndez Guasti, M. and Moya-Cessa, H. M.},
	doi = {10.1088/0305-4470/36/8/305},
	journal = {Journal of Physics A: Mathematical and General},
	number = {8},
	pages = {2069--2076},
	title = {Solution of the Schr{\"o}dinger equation for time-dependent {1D} harmonic oscillators using the orthogonal functions invariant},
	volume = {36},
	year = {2003}}

@book{Folland1989,
	address = {Princeton, NJ},
	author = {Folland, Gerald B.},
	doi = {10.1515/9781400882427},
	publisher = {Princeton University Press},
	series = {Annals of Mathematics Studies},
	title = {Harmonic Analysis in Phase Space},
	volume = {122},
	year = {1989}}

@article{Kimble1998,
	author = {Kimble, H. J.},
	doi = {10.1238/Physica.Topical.076a00127},
	journal = {Physica Scripta},
	pages = {127--137},
	title = {Strong interactions of single atoms and photons in cavity QED},
	volume = {T76},
	year = {1998}}

@article{Purcell1946,
	author = {Purcell, E. M.},
	doi = {10.1103/PhysRev.69.674.2},
	journal = {Physical Review},
	pages = {681},
	title = {Spontaneous Emission Probabilities at Radio Frequencies},
	volume = {69},
	year = {1946}}

@article{SpatMode,
	author = {Morales Rodr\'{\i}guez, M. P. and Herrera, E. Garc\'{\i}a and Maga\~na-Loaiza, O. S. and Perez-Garcia, B. and Guti\'errez, F. Marroqu\'{\i}n and Rodr\'{\i}guez-Lara, B. M.},
	doi = {10.1103/PhysRevA.110.033523},
	issue = {3},
	journal = {Phys. Rev. A},
	month = {Sep},
	numpages = {16},
	pages = {033523},
	publisher = {American Physical Society},
	title = {Spatial-light-mode analogs of generalized quantum coherent states},
	url = {https://link.aps.org/doi/10.1103/PhysRevA.110.033523},
	volume = {110},
	year = {2024}}

@article{Lewis1,
	author = {Lewis, H. R.},
	doi = {10.1103/PhysRevLett.18.510},
	issue = {13},
	journal = {Phys. Rev. Lett.},
	month = {Mar},
	numpages = {0},
	pages = {510--512},
	publisher = {American Physical Society},
	title = {Classical and Quantum Systems with Time-Dependent Harmonic-Oscillator-Type Hamiltonians},
	url = {https://link.aps.org/doi/10.1103/PhysRevLett.18.510},
	volume = {18},
	year = {1967}}

@article{Lewis2,
	author = {Lewis, H. R.},
	doi = {10.1103/PhysRevLett.18.636.2},
	issue = {15},
	journal = {Phys. Rev. Lett.},
	month = {Apr},
	numpages = {0},
	pages = {636--636},
	publisher = {American Physical Society},
	title = {Classical and Quantum Systems with Time-Dependent Harmonic-Oscillator-Type Hamiltonians},
	url = {https://link.aps.org/doi/10.1103/PhysRevLett.18.636.2},
	volume = {18},
	year = {1967}}

@article{Lewis,
	author = {Lewis, H. R. and Riesenfeld, W. B.},
	doi = {10.1063/1.1664991},
	issue = {8},
	journal = {J. Math. Phys.},
	month = {08},
	numpages = {0},
	pages = {1458--1473},
	publisher = {American Institute of Physics},
	title = {An Exact Quantum Theory of the Time-Dependent Harmonic Oscillator and of a Charged Particle in a Time-Dependent Electromagnetic Field},
	volume = {10},
	year = {1969}}

@article{Schattschneider_2014,
	author = {Schattschneider, P. and Schachinger, Th. and St{\"o}ger-Pollach, M. and L{\"o}ffler, S. and Steiger-Thirsfeld, A. and Bliokh, K. Y. and Nori, Franco},
	date = {2014/08/08},
	doi = {10.1038/ncomms5586},
	id = {Schattschneider2014},
	isbn = {2041-1723},
	journal = {Nature Communications},
	number = {1},
	pages = {4586},
	title = {Imaging the dynamics of free-electron Landau states},
	url = {https://doi.org/10.1038/ncomms5586},
	volume = {5},
	year = {2014}}

@article{Ivanov1,
	author = {Ivanov, Igor P. and Korchagin, Nikolai and Pimikov, Alexandr and Zhang, Pengming},
	doi = {10.1103/PhysRevD.101.016007},
	issue = {1},
	journal = {Phys. Rev. D},
	month = {Jan},
	numpages = {13},
	pages = {016007},
	publisher = {American Physical Society},
	title = {Kinematic surprises in twisted-particle collisions},
	url = {https://link.aps.org/doi/10.1103/PhysRevD.101.016007},
	volume = {101},
	year = {2020}}

@article{Ivanov2,
	author = {Ivanov, Igor P. and Korchagin, Nikolai and Pimikov, Alexandr and Zhang, Pengming},
	doi = {10.1103/PhysRevD.101.096010},
	issue = {9},
	journal = {Phys. Rev. D},
	month = {May},
	numpages = {17},
	pages = {096010},
	publisher = {American Physical Society},
	title = {Twisted particle collisions: A new tool for spin physics},
	url = {https://link.aps.org/doi/10.1103/PhysRevD.101.096010},
	volume = {101},
	year = {2020}}

@article{Filina_2023,
	author = {Filina, N. V. and Baturin, S. S.},
	doi = {10.1103/PhysRevA.108.012219},
	issue = {1},
	journal = {Phys. Rev. A},
	month = {Jul},
	numpages = {13},
	pages = {012219},
	publisher = {American Physical Society},
	title = {Unitary equivalence of twisted quantum states},
	url = {https://link.aps.org/doi/10.1103/PhysRevA.108.012219},
	volume = {108},
	year = {2023}}

@article{Bliokh_2017,
	author = {Bliokh, K.Y. and Ivanov, I.P. and Guzzinati, G. and Clark, L. and Van Boxem, R. and B{\'e}ch{\'e}, A. and Juchtmans, R. and Alonso, M.A. and Schattschneider, P. and Nori, F. and Verbeeck, J.},
	doi = {10.1016/j.physrep.2017.05.006},
	issn = {0370-1573},
	journal = {Physics Reports},
	month = may,
	pages = {1--70},
	publisher = {Elsevier BV},
	title = {Theory and applications of free-electron vortex states},
	url = {http://dx.doi.org/10.1016/j.physrep.2017.05.006},
	volume = {690},
	year = {2017}}

@article{Ivanov_2022,
	author = {Ivanov, Igor P.},
	doi = {10.1016/j.ppnp.2022.103987},
	issn = {0146-6410},
	journal = {Progress in Particle and Nuclear Physics},
	month = nov,
	pages = {103987},
	publisher = {Elsevier BV},
	title = {Promises and challenges of high-energy vortex states collisions},
	url = {http://dx.doi.org/10.1016/j.ppnp.2022.103987},
	volume = {127},
	year = {2022}}

@article{Bliokh_2007,
	author = {Bliokh, Konstantin Yu. and Bliokh, Yury P. and Savel'ev, Sergey and Nori, Franco},
	doi = {10.1103/PhysRevLett.99.190404},
	issue = {19},
	journal = {Phys. Rev. Lett.},
	month = {Nov},
	numpages = {4},
	pages = {190404},
	publisher = {American Physical Society},
	title = {Semiclassical Dynamics of Electron Wave Packet States with Phase Vortices},
	url = {https://link.aps.org/doi/10.1103/PhysRevLett.99.190404},
	volume = {99},
	year = {2007}}

@article{Karlovets_2023,
	author = {Karlovets, D. and Di Piazza, A.},
	doi = {10.1103/PhysRevD.108.063007},
	issue = {6},
	journal = {Phys. Rev. D},
	month = {Sep},
	numpages = {14},
	pages = {063007},
	publisher = {American Physical Society},
	title = {Emission of twisted photons by a scalar charged particle in a strong magnetic field},
	url = {https://link.aps.org/doi/10.1103/PhysRevD.108.063007},
	volume = {108},
	year = {2023}}

@article{Dirac_1928,
	author = {Dirac, Paul Adrien Maurice and Fowler, Ralph Howard},
	doi = {10.1098/rspa.1928.0023},
	journal = {Proceedings of the Royal Society of London. Series A, Containing Papers of a Mathematical and Physical Character},
	number = {778},
	pages = {610-624},
	title = {The quantum theory of the electron},
	volume = {117},
	year = {1928}}

@article{Guerrero_2014,
	author = {L{\'o}pez-Ruiz, Francisco and Guerrero, J.},
	doi = {10.1088/1742-6596/538/1/012015},
	journal = {Journal of Physics Conference Series},
	month = {10},
	pages = {012015},
	title = {Generalizations of the Ermakov system through the Quantum Arnold Transformation},
	volume = {538},
	year = {2014}}

@article{Enk_2021,
	author = {Melkani, Abhijeet and van Enk, S. J.},
	doi = {10.1103/PhysRevResearch.3.033060},
	issue = {3},
	journal = {Phys. Rev. Res.},
	month = {Jul},
	numpages = {8},
	pages = {033060},
	publisher = {American Physical Society},
	title = {Electron vortex beams in nonuniform magnetic fields},
	url = {https://link.aps.org/doi/10.1103/PhysRevResearch.3.033060},
	volume = {3},
	year = {2021}}

@article{Aldaya_2011,
	author = {Aldaya, V and Coss{\'\i}o, F and Guerrero, J and L{\'o}pez-Ruiz, F F},
	doi = {10.1088/1751-8113/44/6/065302},
	issn = {1751-8121},
	journal = {Journal of Physics A: Mathematical and Theoretical},
	month = jan,
	number = {6},
	pages = {065302},
	publisher = {IOP Publishing},
	title = {The quantum Arnold transformation},
	url = {http://dx.doi.org/10.1088/1751-8113/44/6/065302},
	volume = {44},
	year = {2011}}

@article{Silenko_2021,
	author = {Zou, Liping and Zhang, Pengming and Silenko, Alexander J.},
	doi = {10.1103/PhysRevA.103.L010201},
	issue = {1},
	journal = {Phys. Rev. A},
	month = {Jan},
	numpages = {7},
	pages = {L010201},
	publisher = {American Physical Society},
	title = {General quantum-mechanical solution for twisted electrons in a uniform magnetic field},
	url = {https://link.aps.org/doi/10.1103/PhysRevA.103.L010201},
	volume = {103},
	year = {2021}}

@article{KarlovetsNJP2021,
	author = {Karlovets, Dmitry},
	doi = {10.1088/1367-2630/abeacc},
	journal = {New Journal of Physics},
	number = {3},
	pages = {033048},
	title = {Vortex particles in axially symmetric fields and applications of the quantum Busch theorem},
	volume = {23},
	year = {2021}}

@article{Filina_2025,
	author = {Filina, N. V. and Baturin, S. S.},
	doi = {10.1103/y1d1-dzqh},
	journal = {Phys. Rev. A},
	month = {Jan},
	pages = {--},
	publisher = {American Physical Society},
	title = {Universal analytic solution for the quantum transport of structured matter waves in magnetic optics},
	url = {https://link.aps.org/doi/10.1103/y1d1-dzqh},
	year = {2026}}

@book{SokolovTernov1974,
	address = {Moscow},
	author = {Sokolov, A. A. and Ternov, I. M.},
	language = {russian},
	note = {In Russian. Bibliography: pp. 380-385. Subject index: pp. 386-391.},
	pages = {391},
	printrun = {4000},
	publisher = {Nauka},
	title = {Relativistic Electron},
	totalpages = {391},
	year = {1974}}

@article{Verbeeck_2010,
	author = {Verbeeck, Jo and Tian, H and Schattschneider, Peter},
	doi = {10.1038/nature09366},
	journal = {Nature},
	month = {09},
	pages = {301-4},
	title = {Production and application of electron vortex beams},
	volume = {467},
	year = {2010}}

@article{doi:10.1126/science.1198804,
	author = {Benjamin J. McMorran and Amit Agrawal and Ian M. Anderson and Andrew A. Herzing and Henri J. Lezec and Jabez J. McClelland and John Unguris},
	doi = {10.1126/science.1198804},
	journal = {Science},
	number = {6014},
	pages = {192-195},
	title = {Electron Vortex Beams with High Quanta of Orbital Angular Momentum},
	volume = {331},
	year = {2011}}

@article{Saitoh_2012,
	author = {Saitoh, Koh and Hasegawa, Yuya and Tanaka, Nobuo and Uchida, Masaya},
	doi = {10.1093/jmicro/dfs036},
	issn = {0022-0744},
	journal = {Journal of Electron Microscopy},
	month = {03},
	number = {3},
	pages = {171-177},
	title = {Production of electron vortex beams carrying large orbital angular momentum using spiral zone plates},
	volume = {61},
	year = {2012}}

@article{Bliokh_2006,
	author = {Bliokh, Konstantin Yu.},
	doi = {10.1103/PhysRevLett.97.043901},
	issue = {4},
	journal = {Phys. Rev. Lett.},
	month = {Jul},
	numpages = {4},
	pages = {043901},
	publisher = {American Physical Society},
	title = {Geometrical Optics of Beams with Vortices: Berry Phase and Orbital Angular Momentum Hall Effect},
	url = {https://link.aps.org/doi/10.1103/PhysRevLett.97.043901},
	volume = {97},
	year = {2006}}

@article{Alexeyev_2006,
	author = {Alexeyev, C N and Yavorsky, M A},
	doi = {10.1088/1464-4258/8/9/008},
	journal = {Journal of Optics A: Pure and Applied Optics},
	month = {jul},
	number = {9},
	pages = {752},
	title = {Topological phase evolving from the orbital angular momentum of `coiled' quantum vortices},
	url = {https://doi.org/10.1088/1464-4258/8/9/008},
	volume = {8},
	year = {2006}}

@article{Bliokh_2014,
	author = {Schattschneider, Peter and Schachinger, Thomas and St{\"o}ger-Pollach, M. and L{\"o}ffler, Stefan and Steiger-Thirsfeld, A and Bliokh, Konstantin and Nori, Franco},
	doi = {10.1038/ncomms5586},
	journal = {Nature communications},
	month = {08},
	pages = {4586},
	title = {Imaging the dynamics of free-electron Landau states},
	volume = {5},
	year = {2014}}

@article{Schachinger_2015,
	author = {T. Schachinger and S. L{\"o}ffler and M. St{\"o}ger-Pollach and P. Schattschneider},
	doi = {https://doi.org/10.1016/j.ultramic.2015.06.004},
	issn = {0304-3991},
	journal = {Ultramicroscopy},
	pages = {17-25},
	title = {Peculiar rotation of electron vortex beams},
	url = {https://www.sciencedirect.com/science/article/pii/S0304399115001382},
	volume = {158},
	year = {2015}}

@article{Wang_2021,
	author = {Wang, Yan and Jia, Chenglong and Zhang, Pengming},
	doi = {10.1063/5.0039479},
	issn = {1077-3118},
	journal = {Applied Physics Letters},
	month = feb,
	number = {8},
	publisher = {AIP Publishing},
	title = {Detection of magnetic impurities using electron vortex beams},
	url = {http://dx.doi.org/10.1063/5.0039479},
	volume = {118},
	year = {2021}}

@article{Cozzolino_2019,
	author = {Cozzolino, Daniele and Bacco, Davide and Da Lio, Beatrice and Ingerslev, Kasper and Ding, Yunhong and Dalgaard, Kjeld and Kristensen, Poul and Galili, Michael and Rottwitt, Karsten and Ramachandran, Siddharth and Oxenl\o{}we, Leif Katsuo},
	doi = {10.1103/PhysRevApplied.11.064058},
	issue = {6},
	journal = {Phys. Rev. Appl.},
	month = {Jun},
	numpages = {12},
	pages = {064058},
	publisher = {American Physical Society},
	title = {Orbital Angular Momentum States Enabling Fiber-based High-dimensional Quantum Communication},
	url = {https://link.aps.org/doi/10.1103/PhysRevApplied.11.064058},
	volume = {11},
	year = {2019}}

@article{Grillo_2017,
	author = {Harvey, Tyler and Venturi, Federico and Pierce, Jordan and Balboni, Roberto and Bouchard, Frederic and Gazzadi, Gian Carlo and Frabboni, Stefano and Tavabi, Amir and Li, Zi-An and Dunin-Borkowski, Rafal and Boyd, Robert and McMorran, Benjamin and Karimi, Ebrahim},
	doi = {10.1038/s41467-017-00829-5},
	journal = {Nature Communications},
	month = {09},
	title = {Observation of nanoscale magnetic fields using twisted electron beams},
	volume = {8},
	year = {2017}}

@article{huge_L,
	author = {Tavabi, A. H. and Rosi, P. and Roncaglia, A. and Rotunno, E. and Beleggia, M. and Lu, P.-H. and Belsito, L. and Pozzi, G. and Frabboni, S. and Tiemeijer, P. and Dunin-Borkowski, R. E. and Grillo, V.},
	doi = {10.1063/5.0093411},
	issn = {0003-6951},
	journal = {Applied Physics Letters},
	month = {08},
	number = {7},
	pages = {073506},
	title = {Generation of electron vortex beams with over 1000 orbital angular momentum quanta using a tunable electrostatic spiral phase plate},
	volume = {121},
	year = {2022}}

@article{Zaytsev_2017,
	author = {Zaytsev, V. A. and Serbo, V. G. and Shabaev, V. M.},
	doi = {10.1103/PhysRevA.95.012702},
	issue = {1},
	journal = {Phys. Rev. A},
	month = {Jan},
	numpages = {8},
	pages = {012702},
	publisher = {American Physical Society},
	title = {Radiative recombination of twisted electrons with bare nuclei: Going beyond the Born approximation},
	url = {https://link.aps.org/doi/10.1103/PhysRevA.95.012702},
	volume = {95},
	year = {2017}}

@article{Filina_2024,
	author = {Filina, N. V. and Baturin, S. S.},
	doi = {10.1103/PhysRevA.110.022204},
	issue = {2},
	journal = {Phys. Rev. A},
	month = {Aug},
	numpages = {8},
	pages = {022204},
	publisher = {American Physical Society},
	title = {Twisted charged particles in the uniform magnetic field with broken symmetry},
	url = {https://link.aps.org/doi/10.1103/PhysRevA.110.022204},
	volume = {110},
	year = {2024}}

@article{Allen_1992,
	author = {Allen, Les and Beijersbergen, Marco and Spreeuw, Robert and Woerdman, J.},
	doi = {10.1103/PhysRevA.45.8185},
	journal = {Physical review. A},
	month = {07},
	pages = {8185-8189},
	title = {Orbital angular momentum of light and transformation of Laguerre Gaussian Laser modes},
	volume = {45},
	year = {1992}}

@article{Foldy_1950,
	author = {Foldy, Leslie L. and Wouthuysen, Siegfried A.},
	doi = {10.1103/PhysRev.78.29},
	issue = {1},
	journal = {Phys. Rev.},
	month = {Apr},
	numpages = {0},
	pages = {29--36},
	publisher = {American Physical Society},
	title = {On the Dirac Theory of Spin 1/2 Particles and Its Non-Relativistic Limit},
	url = {https://link.aps.org/doi/10.1103/PhysRev.78.29},
	volume = {78},
	year = {1950}}

@article{Silenko_2025,
	author = {Meng, Qi and Liu, Xuan and Ma, Wei and Yang, Zhen and Lu, Liang and Silenko, Alexander J. and Zhang, Pengming and Zou, Liping},
	doi = {10.1103/8jrx-p9rz},
	issue = {2},
	journal = {Phys. Rev. Res.},
	month = {Jun},
	numpages = {8},
	pages = {023306},
	publisher = {American Physical Society},
	title = {Generalized Gouy rotation of electron vortex beams in uniform magnetic fields},
	url = {https://link.aps.org/doi/10.1103/8jrx-p9rz},
	volume = {7},
	year = {2025}}

\clearpage
\onecolumngrid

\begin{center}
{\large\bfseries Supplemental Material for
``Geometry-Enabled Radiation from Structured Paraxial Electrons''\par}
\vspace{1em}
M.~S.~Epov, I.~E.~Shenderovich, and S.~S.~Baturin

\vspace{0.75em}
\begin{minipage}{0.9\textwidth}
\small
The Supplemental Material contains the explicit evaluation of the
displacement-operator matrix elements, the photon-wave-vector estimates used
in the paraxial approximation, and the derivation of the emission amplitude
and rate in a locally field-free region.
\end{minipage}
\end{center}

\setcounter{section}{0}

\section{Evaluation of the form factor \label{app:2}}

It is known that
\begin{equation}
  \label{eq:laguerre-01}
  \hat{D}(\alpha)=\exp(\alpha \hat a^\dagger-\alpha^*\hat a)
  =e^{-|\alpha|^2/2}\,e^{\alpha \hat a^\dagger}e^{-\alpha^*\hat a}.
\end{equation}

For the matrix element we write
\begin{align}
  \label{eq:laguerre-02}
  &\langle n_f|\hat D(\alpha)|n_i\rangle
  =e^{-|\alpha|^2/2}\,
  \langle n_f\left|e^{\alpha \hat a^\dagger}e^{-\alpha^*\hat a}\right|n_i\rangle \nonumber\\
  &=e^{-|\alpha|^2/2}\sum_{k=0}^{\infty}\sum_{m=0}^{\infty}
  \frac{\alpha^k(-\alpha^*)^m}{k!\,m!}
  \left\langle n_f\left|(\hat a^\dagger)^k \hat a^m\right|n_i\right\rangle .
\end{align}
We assume $m\le n_i$ and use the standard action of the ladder operators
\begin{align}
  \label{eq:laguerre-03}
  \hat a^m|n\rangle &= \sqrt{\frac{n!}{(n-m)!}}\;|n-m\rangle, \\
  (\hat a^\dagger)^k|n\rangle &= \sqrt{\frac{(n+k)!}{n!}}\;|n+k\rangle .
\end{align}
Applying $\hat a^m$ first and then $(\hat a^\dagger)^k$ yields
\begin{equation}
  \label{eq:laguerre-04}
  \left\langle n_f\left|(\hat a^\dagger)^k \hat a^m\right|n_i\right\rangle
  =
  \frac{\sqrt{n_i!\,n_f!}}{(n_i-m)!}\;
  \delta_{n_f,\;n_i-m+k},
\end{equation}
where implicitly $m\le n_i$ (otherwise the term vanishes). We introduce
\begin{equation}
  \ell \equiv n_f-n_i>0,
\end{equation}
so that the Kronecker delta enforces $k=\ell+m$. Substituting this constraint into
Eq.~\eqref{eq:laguerre-02} and using $\alpha^{\ell+m}(-\alpha^*)^m=\alpha^\ell(-1)^m|\alpha|^{2m}$,
we obtain the single finite sum
\begin{align}
  \label{eq:laguerre-05}
  &\langle n_f|\hat D(\alpha)|n_i\rangle
  = \\
  &e^{-|\alpha|^2/2}\,\alpha^\ell\,\sqrt{n_i!\,n_f!}
  \sum_{m=0}^{n_i}\frac{(-1)^m\,|\alpha|^{2m}}{m!\,(\ell+m)!\,(n_i-m)!}= \nonumber\\
  &
  e^{-|\alpha|^2/2}\,\sqrt{\frac{n_i!}{n_f!}}\,\alpha^\ell
  \sum_{m=0}^{n_i}(-1)^m\frac{n_f!}{(n_i-m)!\,(\ell+m)!\,m!}\,|\alpha|^{2m}. \nonumber
\end{align}
The generalized Laguerre polynomial is defined as
\begin{equation}
  \label{eq:laguerre-06}
  \mathcal{L}_n^{\ell}(x)=\sum_{m=0}^{n}(-1)^m\frac{(n+\ell)!}{(n-m)!\,(\ell+m)!\,m!}\,x^m.
\end{equation}
With $n=n_i$, $\ell=n_f-n_i$, and $(n_i+\ell)!=n_f!$, Eq.~\eqref{eq:laguerre-05} becomes
\begin{align}
  \label{eq:laguerre-07}
  &\langle n_f|\hat D(\alpha)|n_i\rangle
  = \\ \nonumber
  &e^{-|\alpha|^2/2}\,
  \sqrt{\frac{n_i!}{n_f!}}\,
  \alpha^{\,n_f-n_i}\,
  \mathcal{L}_{n_i}^{n_f-n_i}\!\left(|\alpha|^2\right).
\end{align}

This is valid for the case $n_f \geq n_i$. In the case $n_i>n_f$, one similarly obtains
\begin{align}
\label{eq:laguerre-08}
&\langle n_f|\hat D(\alpha)|n_i\rangle
= \\ \nonumber
&e^{-|\alpha|^2/2}\,
\sqrt{\frac{n_f!}{n_i!}}\,
(-\alpha^{*})^{\,n_i-n_f}\,
\mathcal{L}_{n_f}^{n_i-n_f}\!\left(|\alpha|^2\right),
\end{align}

Finally, using Eqs.\eqref{eq:laguerre-07} and \eqref{eq:laguerre-08}, we can write the required form factors.
For the case of $n^f_\sigma\geq n^i_\sigma$ we have
\begin{align}
    &\mathcal{F}_{n^f_\sigma,n^i_\sigma}(\kappa_\sigma)=\\ \nonumber&(-i\kappa_\sigma^*)^{n^f_\sigma-n^i_\sigma}\sqrt{\frac{n^i_\sigma!}{n^f_\sigma!}}\exp\left(-\frac{|\kappa_\sigma|^2}{2} \right)\mathcal{L}_{n^i_\sigma}^{n^f_\sigma-n^i_\sigma}(|\kappa_\sigma|^2),
\end{align}
and for the case of $n^f_\sigma\leq n^i_\sigma$
\begin{align}
    &\mathcal{F}_{n^f_\sigma,n^i_\sigma}(\kappa_\sigma)=\\ \nonumber&(-i\kappa_\sigma)^{n^i_\sigma-n^f_\sigma}\sqrt{\frac{n^f_\sigma!}{n^i_\sigma!}}\exp\left(-\frac{|\kappa_\sigma|^2}{2} \right)\mathcal{L}_{n^f_\sigma}^{n^i_\sigma-n^f_\sigma}(|\kappa_\sigma|^2).
\end{align}

\section{Photon wave vector \label{app:1}}

Since the transverse dynamics is governed by a two-dimensional oscillator, $\hat{\pi}_\perp^{\,2}$ is proportional to the total transverse excitation number $N=n_+ + n_-$ (in unnormalized units, $\Delta(\hat{\pi}_\perp^{\,2})\propto \Delta N/\rho_H^2$). In the paraxial regime, the energy change is dominated by the change of transverse kinetic energy, thus we get
\begin{align}
\omega(z)&\simeq E_i-E_f \simeq\frac{\Delta (\hat{\pi}_\perp^2)}{2E} =\nonumber \\ &=\frac{\beta}{2k\rho_H^2}\left[\left(b'^2+\frac{1}{b^2}+\Omega^2 b^2 \right)\Delta N+2\Omega \Delta l\right],
\end{align}
where $\beta=k/E$ is the relativistic $\beta$-factor. Thus one obtains
\begin{align}
\left(\frac{k_z^{\omega}}{k}\right)\chi^2&\simeq \\ \nonumber
&\frac{\beta}{2}\cos\theta \left[\left(b'^2+\frac{1}{b^2}+\Omega^2 b^2 \right)\Delta N+2\Omega \Delta l\right].
\end{align}
Similarly, one obtains
\begin{align}
\rho_H k_\perp^{\omega} &\simeq \\ \nonumber &\frac{\beta\sin\theta}{2\chi} \left[\left(b'^2+\frac{1}{b^2}+\Omega^2 b^2 \right)\Delta N+2\Omega \Delta l\right].
\end{align}

\section{Field-free region \label{app:3}}

In a field-free region, $\Omega=0$ and the Ermakov equation reduces to
\begin{align}
    b''=\frac{1}{b^3}.
\end{align}
Assuming the beam waist is located exactly at the center of the region ($b'(0)=0$), we obtain the well-known envelope
\begin{align}
    b(z)=b_0\sqrt{1+\frac{z^2}{b_0^4}}
\end{align}
Restoring dimensions via $z=k\rho_H^2 z_{ph}$, we obtain
\begin{align}
    \mathrm{w}(z)=\mathrm{w}_0\sqrt{1+\left(\frac{z_{ph}}{z_R}\right)^2}.
\end{align}
Here $\mathrm{w}(z)$ is the state width, $\mathrm{w}_0=\rho_H b_0$ is the state waist, and $z_R=k\rho_H^2 b_0^2$ is the Rayleigh length.
We have the following identity
\begin{align}
\label{eq:invA}
    b'(z)^2+\frac{1}{b(z)^2}=\frac{1}{b_0^2}.
\end{align}
Using Eq.\eqref{eq:invA}, we get
\begin{align}
\label{eq:omsim}
    \omega\simeq \frac{\beta \Delta N}{2 b_0^2 k \rho_H^2}
\end{align}
and 
\begin{align}
\left(\frac{k_z^{\omega}}{k}\right)\chi^2\simeq \frac{\beta \Delta N}{2 b_0^2}\cos\theta.
\end{align}
as well as
\begin{align}
\rho_H k_\perp^{\omega} \simeq \frac{\beta\Delta N\sin\theta}{2\chi b_0^2}.
\end{align}
We neglect recoil and estimate $\Delta k$ as 
\begin{align}
    \frac{\Delta k}{k}\chi^2 \simeq \frac{\Delta N}{2 b_0^2}
\end{align}
The $C$ coefficients simplify to
\begin{align}
\label{eq:cfuncFS}
              &C_{a,\sigma}=\frac{1}{b_0}\exp\left[-i \arctan\left( \frac{b_0^2}{z}\right) \right], \nonumber\\
    &C_{a^\dagger,\sigma}=\frac{1}{b_0}\exp\left[i \arctan\left( \frac{b_0^2}{z}\right) \right].
\end{align}
The Lewis phase simplifies to
\begin{align}
    \int\limits_0^z \frac{d\bar{z}}{b(\bar{z})^2}=\arctan\left[\frac{z}{b_0^2} \right]
\end{align}
In the dipole case, $\Delta N=N^i-N^f=1$ (for $\ell^i>0$), $\Delta \ell=\ell^i-\ell^f=1$, and for $n_r=0$ (so that $n_+=\ell$ and $n_-=0$) the factor $\mathcal{P}(z)$ is
\begin{align}
    &\mathcal{P}_{n^f_+ n^i_+}(z)=\frac{1}{b_0}
    \exp\left[-i \arctan \left(\frac{b_0^2}{z} \right)\right]\sqrt{\ell^i} 
\end{align}
With $\mathcal{P}$ given above, we can evaluate the $z$ integral
\begin{align}
    &\left|\int\limits_{-L/2}^{L/2} s_\perp(\bar{z})d\bar{z}\right|^2=\frac{\ell^i}{b_0^2}\frac{(1-\lambda\cos\theta)^2}{8}\times \\
    &\left|\int\limits_{-L/2}^{0}\exp\left[iA \bar{z}+i\frac{\pi}{2}\right]d\bar{z}+
    \int\limits_{0}^{L/2} \exp\left[iA \bar{z}-i\frac{\pi}{2}\right] d\bar{z}\right|^2 \nonumber\\ 
    &=\frac{\ell^i}{b_0^2}\frac{(1-\lambda\cos\theta)^2}{8}\frac{L^2\left[\sin X \right]^4}{X^2}\nonumber
\end{align}
Above
\begin{align}
    A=\frac{1-\beta\cos\theta}{2b_0^2},
\end{align}
and
\begin{align}
    X=\frac{1-\beta \cos\theta}{8b_0^2}L
\end{align}
After integrating over $\varphi$ and $\omega$, the rate reads
\begin{align}
    d\dot{{w}}=\alpha\frac{\beta^2}{\chi^2 L^2} \Delta E \sin\theta\left|\int\limits_{-L/2}^{L/2} s_\perp(z) dz\right|^2 d\theta
\end{align}
where $\Delta E$ is given by Eq.~\eqref{eq:omsim}, since the delta function enforces $\omega=\Delta E$.

After summing over the photon polarizations and simplifying, we obtain
\begin{align}
    \frac{d\dot{{w}}}{d\theta}=\ell^i\frac{\alpha\beta^4\gamma m_e}{\chi^4} \frac{\left[\sin X\right]^4}{X^2}\frac{\left[1+(\cos\theta)^2 \right]\sin\theta}{8b_0^4}.
\end{align}

If the beam waist is not centered in the detector region, it is useful to distinguish the waist data $(b_0,z_w)$ from the local detector-center data $(b_c,b'_c)$, where
\begin{align}
b_c=b(0),
\qquad
b'_c=b'(0).
\end{align}
The field-free solution can be written as
\begin{align}
b(z)=\sqrt{b_0^2+\frac{(z-z_w)^2}{b_0^2}}.
\end{align}
Equivalently,
\begin{align}
b_c^2=b_0^2+\frac{z_w^2}{b_0^2},
\qquad
b'_c=-\frac{z_w}{b_0^2 b_c}.
\end{align}
The corresponding squared integral evaluates to
\begin{align}
&\left|\int\limits_{-L/2}^{L/2} s_\perp(\bar{z})d\bar{z}\right|^2
=
\frac{\ell^i}{b_0^2}
\frac{(1-\lambda\cos\theta)^2}{8}
\frac{4}{A^2}
\nonumber\\
&\times
\left[
1
-2\cos(Az_w)\cos\left(\frac{AL}{2}\right)
+\cos^2\left(\frac{AL}{2}\right)
\right],
\end{align}
where
\begin{align}
A=\frac{1-\beta\cos\theta}{2b_0^2}.
\end{align}
When the waist is centered ($z_w=0$), this recovers the $\propto \sin^4(X)/X^2$ dependence derived above. This generalized expression explicitly demonstrates that the radiation in the field-free region is an interference effect critically dependent on the position of the waist relative to the finite observation boundaries. It is important to distinguish this mechanism from ordinary transition or diffraction radiation. While the latter typically require material boundaries or apertures to provide the necessary momentum transfer, the radiation described here arises purely from the internal ``geometric dressing" of the electron state. The effective driving force is the intrinsic wavefront curvature fixed by the Ermakov dynamics, and the emission is modulated by the Gouy phase evolution. Thus, the interaction length $L$ does not act as a physical radiator but as a temporal window that samples the non-stationary geometric phase of the structured electron.

\section{Limiting transitions}
Below we demonstrate limiting transitions to the known cases of a plane wave and a Gaussian paraxial state.
First we take the plane-wave limit of the field-free rate, which corresponds to $b_0\to\infty$. Taking this limit, we find
\begin{align}
    \lim\limits_{b_0\to\infty} R(\theta,L,b_0)\propto \lim\limits_{b_0\to\infty} \frac{1}{b_0^8}=0.
\end{align}

The Gaussian limit is recovered by setting $l^i=0$, which immediately gives $d\dot w/d\theta=0$. 

Finally, in the limit $L\to\infty$ the rate scales as
\begin{align}
    \lim\limits_{L\to \infty} \frac{d\dot{w}}{d\theta}\propto \lim\limits_{L\to \infty} \frac{1}{L^2}=0,
\end{align}
which highlights that the transition is a boundary (edge) effect.

\end{document}